# HDRL Strategy Meeting 2026 Report

Jun 23, 2026

Brian Thomas[1], Robert Candey[1], Jack Ireland[1], Lan Jian[1], Rebecca Ringuette[1,2], and Tressa Helvey-Kalsulke[1,3]

1. Goddard Space Flight Center/NASA Code 672, Greenbelt MD
2. University of Maryland Baltimore County, MD
3. ASRC Federal Systems, MD

## Executive Summary

HDRL (Heliophysics Digital Resource Library, https://hdrl.gsfc.nasa.gov) is a federation of three member organizations: the Heliophysics Data and Model Consortium (HDMC), the Solar Data Analysis Center (SDAC), and the Space Physics Data Facility (SPDF). HDRL's mission is driven and guided by its stakeholders, which include the heliophysics research community, NASA mission scientists, NASA HPD, educators, students and HDRL technical workers.

In order to better engage with this last stakeholder group, HDRL held its first HDRL-wide strategy meeting in January 2026, pulling in a majority of HDRL member staff and selected HDRL grantees. Just over 40 people attended with a few more online to listen to over 30 lightning talks presented by their colleagues and leaders, participate in several breakout and group discussions, and cast 324 votes on over 50 strategic topic ideas contributed by the attendees.

The many ideas generated by attendees can be grouped into seven themes which describe the overall breadth of discussion at the meeting. These themes included:

1. Metadata, Standards, and Semantics
2. Discovery, Search, and Recommendation
3. Modernizing HDRL Users' experience
4. Modernizing HDRL Staff's experience
5. Modernized, Scalable, and Dependable Infrastructure for Compute and Storage
6. Metrics and Impact Assessment
7. Community Engagement and Sustainability

The plethora of topic ideas associated within these themes can be further reduced to 11 prioritized ideas. Considering the gathered information and the interconnections between the prioritized ideas we conclude the following strategic areas for action exist:

- Improve the mission data delivery process to HDRL;

- Formulate and incorporate formal feedback methods into interactions with HDRL stakeholders, both automated and manual;
- Develop and implement a balanced outreach and engagement plan, including HDRL Workshops;
- Investigate means to better capture useful metrics related to the HDRL mission;
- Streamline and modernize HDRL's operations and services, incorporating AI/ML approaches where beneficial;
- Prepare and develop solutions for serving big data, including cloud and on-site HEC[1] solutions;
- Engineer a semantic framework for Heliophysics resource interlinking, working towards a HDRL data knowledge graph; and
- Address gaps in Heliophysics data environment standards through IHDEA[2], the international standards body for Heliophysics data environments, in particular for heliophysics events and semantic resources.

All three of the primary goals of the meeting were successfully met as follows. Staff awareness of HDRL and related activities was achieved via the many presentations and discussions which followed. Meeting attendees were also (re-)exposed to the drivers of HDRL strategy and participated in several exercises to consider gaps in the strategy and potential solutions. Finally the copious output from the meeting attendees active participation also indicates significant engagement and success with the final goal to leverage HDRL staff and grantee expertise to better understand HDRL strategic gaps and solutions. Due to the positive reception and by attendee request, HDRL will plan an improved repeat of the meeting in the next two to three years.

---

[1] High End Compute
[2] International Heliophysics Data Environment Alliance, https://ihdea.net/

## Introduction

HDRL (Heliophysics Digital Resource Library, https://hdrl.gsfc.nasa.gov) is a federation of three member organizations: the Heliophysics Data and Model Consortium (HDMC), the Solar Data Analysis Center (SDAC), and the Space Physics Data Facility (SPDF). Figure 1 provides a schematic of these HDRL member organizations' relationship to one another.

This organizational structure helps the HDRL members better collaborate and coordinate their resources which results in more efficient delivery of support for data-driven research in Heliophysics. There are always a number of pressing choices in what and how this support is delivered and HDRL strategy helps to determine the priorities for this support. HDRL management believes that an important way to guide HDRL strategy is through stakeholder engagement. While obvious stakeholders include the heliophysics research community, NASA HelioPhysics Division (HPD) management, and US and international agency partners, it also includes the workers of HDRL (staff and grantees). This meeting was held specifically to better understand the unique viewpoint of this group of stakeholders.

**Figure 1. Schematic relationship between HDMC, SDAC and SPDF**

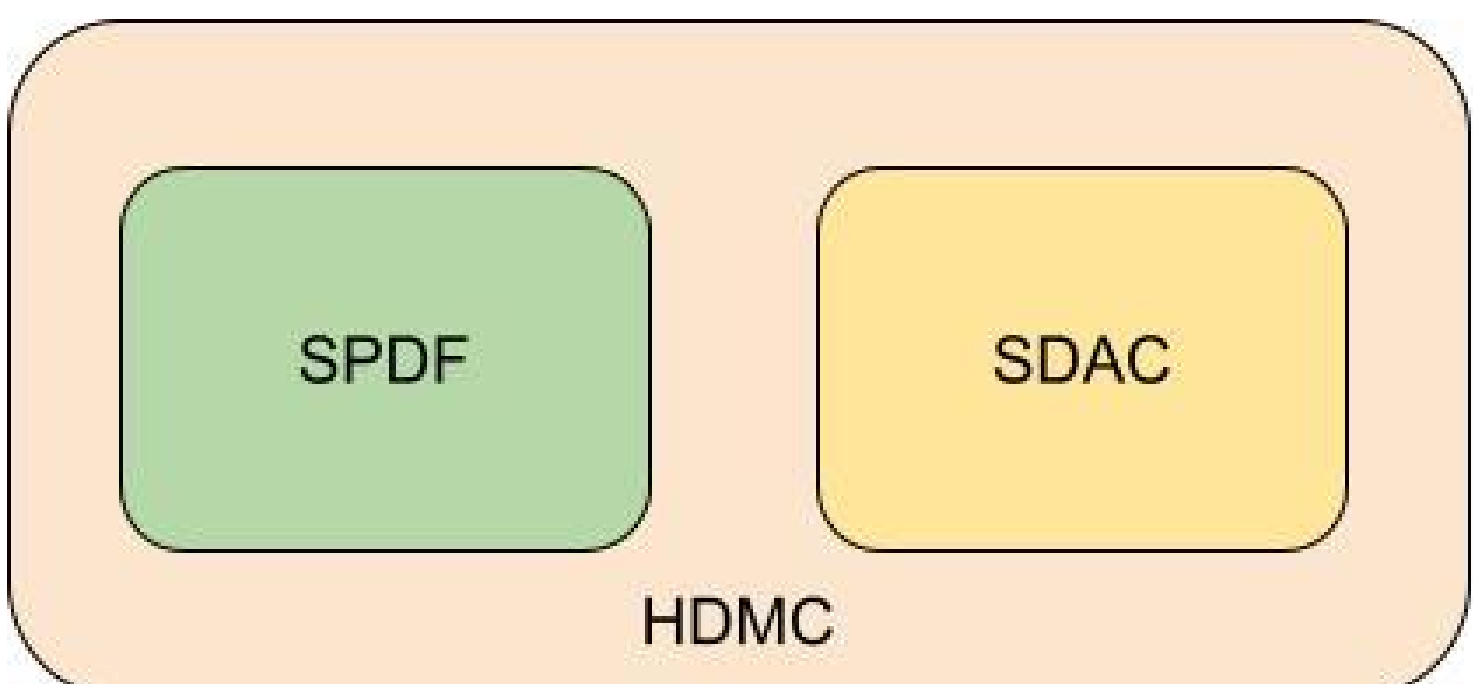

***Figure 1. Schematic relationship between HDMC, SDAC and SPDF.*** *SDAC and SPDF respectively provide solar- and space-physics specific data management, curation, and services for their respective archived data. HDMC provides services and capabilities which help to sew together solar- and space-physics resources for the community.*

The HDRL Strategy Meeting was held Jan 21-22, 2026, with participants drawn from various HDRL member staff and selected HDRL grantees in order to achieve a variety of goals (see Figure 2 for a picture of some of the meeting attendees).

Meeting goals included:

1. Raising staff awareness of HDRL and related activities,
2. Introducing HDRL staff to drivers and strategic concerns which affect HDRL, and
3. Leveraging HDRL staff ingenuity to strategize about current and future challenges.

In this document we describe the structure and content of the meeting and report on the results.

**Figure 2. HDRL Strategy Meeting Attendees**

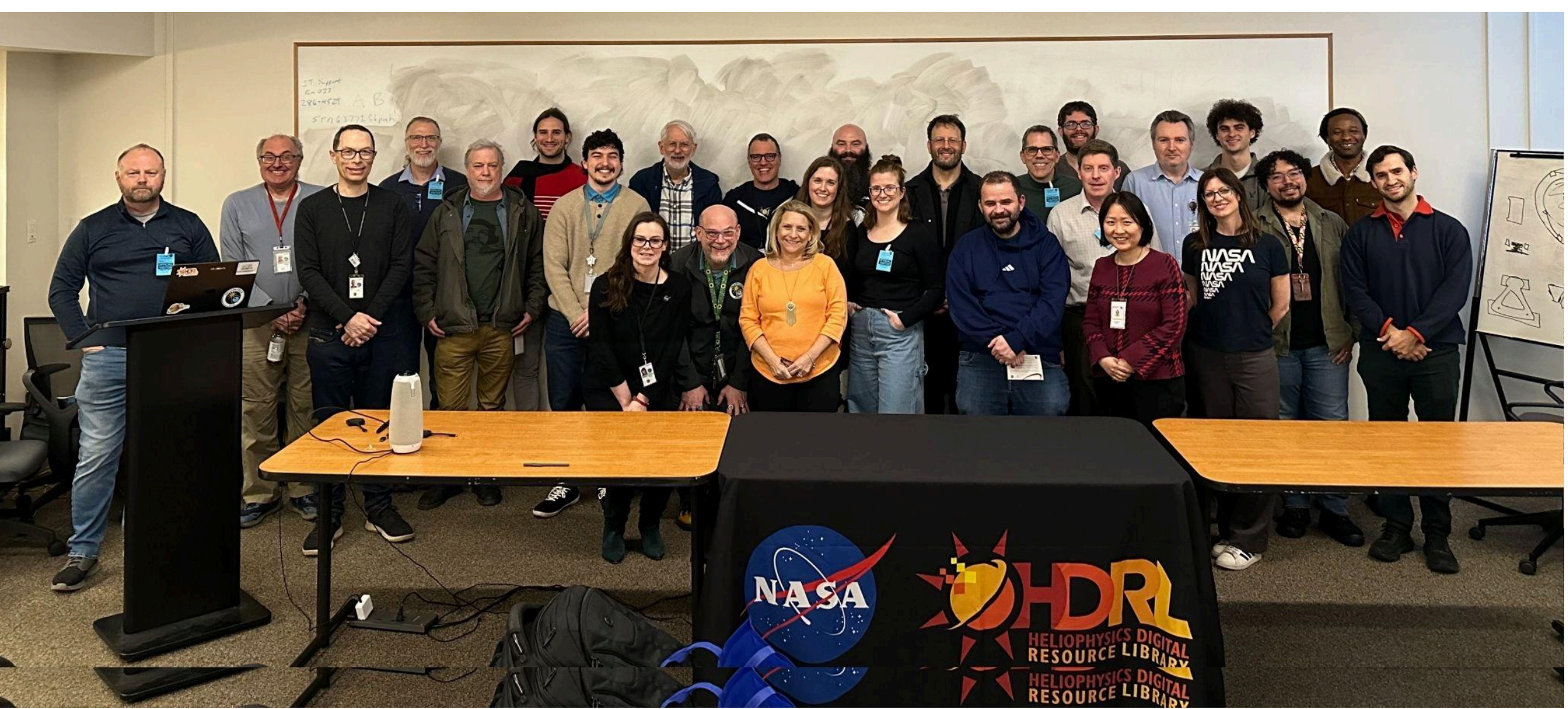

***Figure 2. HDRL Strategy Meeting Attendees.*** A partial group of the meeting attendees photographed on the second day. L-R: Alisdair Davey (NSO), Scott Bordsen (UMBC), Jack Ireland (NASA), Jon Vandegriff (JHUAPL), Jeff Bradford (ADNET), Fernando Morales (UMBC), Zach Boquet (ADNET), Sarah Rourke (ADNET), Bobby Candey (NASA), Alex Young (NASA), Brian Thomas (NASA), Tressa Helvey-Kasulke (ASRC), Rebecca Ringuette (ADNET), Eric Grimes (ADNET), Julie Barnum (LASP), Bob Weigel (GMU), Necdet Percinel (Telophase Corp), Sandy Antunes (JHUAPL), Chris Bard (NASA), Ryan Timmons (Lockheed Martin), Lan Jian (NASA), Andriy Koval (UMBC), Rose Brunning (KBR), Aiden Scharnikow (UMBC), Daniel Garcia Briseno (ADNET), Wale Jaiyeola (Aurora Engineering), and Anthony Buonomo (UMBC).

## Meeting Organization

The two-day meeting was hybrid, and the meeting site was Goddard Space Flight Center (GSFC) Building 21. 47 individuals registered (42 in person, 5 virtual) for the meeting, and roughly that same number attended. We summarize the meeting sessions and related activities below. An agenda with details about presentations and speakers appears in Appendix A.

### Day 1

*HDRL Activities Sessions*

The first day of the meeting focused on the first two goals with over 30 lightning talks given by a large variety of HDRL staff and grantees. These talks began with lightning talks from HDRL leadership, followed by a quick icebreaker exercise on Padlet. The numerous lightning talks that followed were loosely grouped into three sessions around the following areas: “Current Infrastructure”, “Community Connections and Operations”, and “Metadata and Data Management”. Each presenter was instructed to focus on a few high-level objectives for their lightning talk:

1. Define your activity, area, or project, including the part of HDRL you work under.
2. Identify who works with you.
3. Identify the major areas of activity that the team does, limiting to high level details only.
4. Talk about future goals and activities and include one or more challenges. ‘Challenges’ may be broadly defined as things such as pain points like long, expensive, or manual processes or difficult technical problems.

During the presentations, the audience was instructed to contribute ideas related to these talks using the Padlet[3], and it was suggested that this feedback focus on capturing related challenges and/or how to streamline or improve them.

*HDRL Driver Session*

After these three sessions the meeting pivoted to a session on HDRL drivers. A short presentation was then given to define and identify sources of drivers which included a brief discussion of various programmatic/mission, policy, and community drivers. This presentation also served as the launching point for the first breakout discussion of the meeting, focusing on brainstorming topic ideas which would be discussed on the next day of the meeting. Attendees utilized the Padlet to submit their topic ideas and were encouraged to ask and answer questions through the commenting feature. Ideas that belonged to a given type of driver (see Thomas & Ringuette 2026) were loosely categorized.

[3] https://padlet.com/rebeccaringuettenasa/HDRLStrategy2026

### Day 2

*Identification of HDRL Strategic Challenges Sessions*

The second day began with a dedicated time for attendees to rate these ideas on the Padlet. These ratings and the driver categorizations were then used to prioritize the discussion topics for three breakout sessions the same day based on the average rating of the topic multiplied by the number of votes it received. The full list of the topic ideas, their descriptions, and ratings is available in Appendix B.

For each of the three breakout sessions, the moderators chose 1-3 topic ideas to discuss as time allowed. For each topic chosen, the breakout groups were asked to consider the following:

- Improve the description of the challenge as needed.
- What would this affect or ideally improve if solved?
- How would we get started? Who or what skills are needed?
- Make a pitch to get others onboard.

After each breakout session, all groups reconvened together in what was called “report outs” and shared their answers to these questions with the larger group. Some time was then given to discuss any high-level items with the entire group. Notes from the breakout discussion and the report outs were recorded in the “Breakout Notes” google doc linked from the agenda.

At the end of the trio of breakout sessions, a lengthy full group discussion was held to give space for group reflection on the topics discussed. Notes from this discussion are located in the same “Breakout Notes” document on the “Plenary Notes” tab.

## Data Collection

There were 53 separate topic ideas submitted by staff to the Padlet which also included substantial comments and feedback for most of the submitted topic ideas.

In order to filter the list for focused discussion, staff were asked to vote on idea topics using 1-5 ‘stars’. A rating of zero stars was not possible in the interface. These 324 ratings were averaged, multiplied by the number of ratings each topic idea received to calculate an “interest score”, and then sorted to uncover the most interesting topics for discussion during the meeting (see Appendix B for full list). The average number of votes per idea was 6.11 +/- 2.76 with an average rating of 3.87 +/- 1.00 stars (errors equal to one standard deviation). The average interest score across the entire set of topic ideas was 24.08 +/- 13.41.

The groups were free to pick from the list which ideas to explore. As a result, some topic ideas with high interest scores were not discussed, and some below the average interest score were selected. Furthermore there was a general "SPASE[4]" discussion breakout group in the last session based on multiple requests which was very well attended (more than 10 participants) and covered several topic ideas related to SPASE with a variety of interest scores. The topics from the Padlet discussed during the breakout sessions as chosen by the groups are listed below with a summary of each discussion.

### Breakout Sessions: Topic Ideas Discussed

This section collates the topic ideas which were discussed by breakout groups in the order of their ranking by interest score. Each sub-section includes a summarization of the discussion by the groups.

#### *Put Together a CME Tracker*

This topic idea was the highest ranked by interest score. The "putting together a CME tracker" was discussed as one example of how to test the combined usefulness of HDRL services and was selected for discussion in both a practice and a normal breakout discussion session. The group acknowledged that we (HDRL) work hard to make high-quality, trustworthy services on top of the data we host to better serve the community; however, there is no current workflow to test how smoothly those services can be used together. Example workflows are needed based on those used in publications, preferably complemented by community feedback and involvement. Using CMEs as a starting example was well supported by attendees. The pitch given on this topic idea was "Find the Gap" in HDRL, meaning use this example workflow approach to find the gap in our services and then fix those gaps to make research easier.

#### *Formal Feedback*

This topic idea was ranked second by its interest score. Two breakout groups selected this idea for discussion at the same time indicating the large interest in this area. Identified challenges in this area included the community's hesitancy to use GitHub, the inability of the general public to search for and read content on HelioNauts.org, low attendance at previous HDRL-specific outreach events, and the innate tendency of users to vote "with their feet" instead of submitting complaints. A multi-pronged approach was discussed with the following potential components:

- Automated feedback through HDRL's services, such as a "server temporarily down for maintenance" message automatically sent through the service API during maintenance times when a user attempts to retrieve data and the attempt fails. This method should also “automagically” send notifications to the proper HDRL staff of the failures.

[4] Space Physics Archive Search and Extract metadata format, https://spase-group.org

- Proactive service status reporting, such as the automatically generated reports on the VSO for uptime/downtime/status per source.

- Proactive outreach with a “go to them” approach rather than a “come to us”. One popular example was HDRL engaging with mission-specific meetings in our outreach plan.

- Visually prominent suggestion boxes/forms on relevant websites, including those for services.

- Typical user-support types of venues, such as office hours or regular meetings (possibly supported by a user-focused HDRL Working Group), a public searchable chat channel, and possibly an AI tool to help users diagnose problems.

- Dedicated user workshops where the attendees’ travel expenses are paid by HDRL in return for detailed user feedback on the selected HDRL services in person following usability testing guidelines, ideally supported by trained usability testing personnel.

These efforts should be preceded by some basic structuring for the feedback, such as a template for feedback and predetermined categories for the feedback (e.g., “feature request”, “user question”, and “other”), and some outlining of best practices for obtaining good feedback (e.g., good questions to ask, the typical steps, etc).

*HDRL Workshops*

Ranked third by interest score, this topic idea was also discussed simultaneously in two groups. The discussion was interwoven with the previous two topic ideas, combining ideas about formal feedback and “Bring HDRL to the people” and with sewing HDRL’s services together through example workflows such as the CME tracker, or “Find the Gap” in HDRL. Short 2-hr virtual or hybrid monthly meetings were described as a good starting point, where the content would be a 1-hr how-to tutorial on either how to use a HDRL service or an example phenomenon-focused analysis using HDRL services. The tutorial would be followed by 1 hour of practice with some time for feedback. The tutorial part of the meetings would be recorded and added to a dedicated channel on YouTube (or similar), which would then need to be linked from relevant HDRL websites. These short meetings would build up to potentially longer workshops with more variety, which would be held in conjunction with and in support of existing meetings, including mission-specific meetings and more general science meetings such as CEDAR, GEM, and SHINE. Eventually, a longer workshop could be held similar to PyHC’s summer school but focused on analyses using HDRL services and the supporting service tutorials.

*Build a Knowledge Graph of HDRL Data*

Ranked sixth by interest score, this topic idea was discussed and thought to have multiple uses. A few example uses discussed were

- Suggesting similar datasets based on overlapping metadata, including older mission data, including metadata from publications alongside datasets;
- Suggesting meetings to attend to find subject matter experts related to the dataset;
- Connect data to people, further democratizing access to the relevant experts, whether by email, chat channel on HelioNauts, or other avenues;
- Connect the user to the useful starting-point publications (e.g., reviews, publications from the dataset or instrument team, highly cited historical papers, etc) related to the dataset or topic searched for to aid new researchers; and
- Understanding data interrelationships: for example what data are important to or the foundation of other data, whether from HDRL's perspective or from a funder's perspective.

Some important components noted in the discussion included a reliance on the Science Explorer and internal efforts to connect research publications (and consequently their authors) with datasets, the challenge of establishing and maintaining the relevant ontologies (e.g., phenomena), and the importance of a central funding model for the work to persist. The implementation style discussed was one of a text-based search interface, possibly a chat-bot style, where the user would ask the question and the software would navigate the knowledge graph to find relevant answers. The breakout group participants specifically noted this idea as providing support for a few drivers, specifically the need to promote archival research and the side-by-side use of ground and space data in research, although the participants also highly regarded the connections between people and the research artifacts as a democratizing feature.

*Enhance or Draft SPASE Records from Documents using AI*

A short discussion addressed this topic idea, which was ranked 25th by interest score. A tool based on this topic idea was considered useful for drafting SPASE records, with full automation considered likely acceptable for low priority datasets, such as low level datasets, legacy or backlogged datasets, and researcher-contributed datasets. Regardless of the level of automation, this tool could be useful for accelerating the creation of richer SPASE records, and for making automatic suggestions to update existing SPASE records. The drafted metadata could in some cases be considered as additional entries alongside existing metadata where a SPASE record already exists, such as adding keyword-type fields based on metadata extracted from data documentation or a data management plan. Other existing sources of information for drafting SPASE records were discussed, including HAPI[5] (Weigel et al. 2021), CDAWeb[6], Zenodo[7], and the SKTEditor[8] (Grimes 2025). However, those sources only apply to datasets covered by those services.

*Metrics*

One breakout group chose to combine a few low-ranked metrics-related topic ideas into one discussion, indicating the value and interrelated nature of the collection of metrics-releated topic ideas in aggregate. The included topic ideas were “Build a Metrics System for our Repositories” (ranked 13th by interest score), “Build a Metrics System for Datasets” (ranked 27th), “How to Determine what Data Supported Scientific Breakthroughs?” (ranked 50th). The discussion gave a few examples of metrics to include in such a system, such as dataset access numbers, a measurement of HDRL engagement with users at in-person events, citations in publications to datasets hosted by HDRL. The metrics gathered would be used by HDRL internally, by NASA HQ, and by missions for assessing impact of the datasets produced by a mission or to provide supporting justification for directing efforts. However, it was clearly noted that high metrics for a given dataset is *not* a measurement of scientific impact since the impact of a given dataset on the field is often not clear for several years, even decades.

*How Do We Prepare to Serve Big Data?*

This topic idea was ranked 31st out of over 50 topic ideas by interest score, but was still selected by a breakout group for discussion. DKIST was regarded in the conversation as an example of success in serving large data using Globus but without full computational support. Some discussed components of a potential solution include:

- Incorporating Globus into HDRL services, looking to DKIST and the related implementation in SunPy as successful examples;
- Selecting useful “levels” of data to save instead of saving all levels of data;
- Saving only the latest version of the data;
- Injecting HDRL much earlier into the mission data pipeline, potentially hosting the data processing to reduce the data volumes needed for transfer;
- Computing capability next to the data, possibly hosting the data on HPC resources to be used as a less expensive “cloud” computing platform.

*SPASE*

This breakout session was different from the others in that it did not focus on a particular topic idea, but instead focused on a general topic involving the SPASE metadata format. The discussion was broader in scope  than the other breakout sessions, touching on several related items including various metadata mappings to and from SPASE, reiterating the need for a redesign of SPASE (e.g., SPASE 3.0), the need for automation between SPASE and related metadata systems to keep metadata up-to-date, a simplified metadata submission process, and the desire to incorporate AI tooling to more quickly draft SPASE metadata. The six related topic ideas are:

- “Discuss what needs to be done to finish ISTP/ HAPI/ SPASE mapping”
- “Use vectorization for keywords”
- “Create a SPASELite editor for NASA”
- “Draft SPASE records from SDAC/SPDF metadata using AI”
- “Draft SPASE records from documents using AI”
- “SPASE Statistics Chat Bot”

[5] https://hapi-server.org/
[6] https://cdaweb.gsfc.nasa.gov/
[7] https://zenodo.org/
[8] https://spdf.gsfc.nasa.gov/skteditor/

Overall, the sentiment of the discussion was confidence that HDRL is heading in a good direction with the current array of SPASE-related efforts, but significant increases to metadata quantity and SPASE record drafting speed were possible if AI tooling were developed and included into operations.

## Analysis and Synthesis

There is a substantial amount of data generated during the meeting ranging from the contributed topic ideas, the comments on those same topic ideas and the captured breakout group discussions (see prior section). All of these materials may be considered the ‘content’ of the meeting. This section organizes that content into themes and presents statistical information from the voting.

### Topic Idea Themes

As a first organizing step to get a better understanding of the nature of the meeting content, we aimed to create thematic groupings with fairly equal representation of the topic ideas submitted in each theme. We first passed the entire collection of topic ideas and their respective descriptive text to ChatGPT[9] and asked for these to be grouped thematically. We then reviewed this output for accuracy, split an over-represented group apart, altered the AI-generated groups to remove overlap and inconsistencies, and then mapped each topic idea to the relevant themes. We found a few topic ideas that were substantially related to more than one theme (13), but decided to place these only in the two most related themes rather than in all of the possible themes.

[9] OpenAI. (2023). *ChatGPT* (GPT-5.2, Feb 2025) [Large language model]. https://chat.openai.com

The following themes emerged:

*1. Metadata, Standards, and Semantics*

*How (meta)data are described, structured, and made interoperable*.

This theme includes:
- SPASE / ISTP / HAPI / netCDF4 / HDF5 mappings and standard updates
- DataCite field richness and metadata quality plans
- Schema.org, Croissant, metadata relevant for search engine optimization
- Defining classes of data and tagging
- Standard formats for event lists, units/coordinate frames, and machine learning
- Warning labels, uncertainties, and instrument-specific metadata
- Building and stewarding the heliophysics knowledge commons
- Events standard for both space- and solar-physics

*2. Discovery, Search, and Recommendation*

*Helping users and machines find relevant data.*

This theme includes:
- Dataset and event-based recommendation engines
- Helioviewer as a discovery hub (“holy grail” vision)
- AI-assisted discovery (LLMs over metadata, NL queries)
- SPASE summarization for AI search tools
- Metadata strategies for AI agents and search engines
- Interlinking research artifacts
- Making datasets easier to find across missions and domains

*3. Modernizing HDRL Users’ experience*

*Reducing friction for scientists.*

This theme includes:
- “Submit” button for artifact ingestion
- Community resource / featured tools pages
- Chatbots (service problem reporting, data documentation)
- Easier access to TAP, HEK, and external services
- SPASE creation, including related AI tooling
- Automated user feedback

*4. Modernizing HDRL Staff's experience*

*Reducing friction for staff.*

This theme includes:

- Chatbots (SPASE stats, service problem reporting)
- Rapid first-pass ingestion of event lists
- SPASE (creation, validation, statistics), including related AI tooling
- Automated staff notifications

*5. Modernized, Scalable, and Dependable Infrastructure for Compute and Storage*

*Making sure things work at today's and tomorrow's data volumes and compute needs.*

This theme includes:

- Cloud vs HPC strategy (HelioCloud, HySDS, Globus)
- Hosting, serving and analyzing PB-scale datasets
- Big-data failure modes (e.g., HAPI silently failing)
- Sewing services together (e.g., CME trackers, notebooks on the cloud)
- AI-assisted data access (LLMs / MCP servers over services)
- Copying PB-scale externally hosted NASA datasets into HDRL
- Better file flow and validation into HDRL
- Real-time monitoring dashboards for measuring service maturity and reliability

*6. Metrics and Impact Assessment*

*Knowing what's being used, how, and why it matters.*

This theme includes:

- Repository and dataset-level metrics systems, including access metrics
- Understanding and filtering bot traffic
- Open Science compliance reporting
- Tracking data use in scientific breakthroughs
- Impact metrics derived from knowledge graphs

*7. Community Engagement and Sustainability*

*The human and institutional layer encompassing governance, sustainability, outreach and collaboration.*

This theme includes:

- Formal feedback collection and review mechanisms

- HDRL[10] recommendations process and governance
- Save at-risk datasets
- Funding sustainability and continuation challenges
- Mission–community tool integration
- Workshops, training, and capacity building
- Cross-division partnerships (Earth Science, Sci-X, etc.)

### Statistics

To further distill the meeting content, to check if derived themes sufficiently represent the content of the meeting, and to better understand staff interest and momentum around topic ideas and themes we attempted to analyze the topic ideas and related themes based on their statistical properties.

*Topic Idea - Theme Associations and Ratings*

As previously noted we associated the topic ideas with one or more themes, which may be seen in table B1. Of the 54 topic ideas generated in the meeting, our analysis found that 13 of them overlapped with another theme resulting in a second associated theme. This brings the number of theme-topic idea associations to 66 with an average of about 9 topic ideas associated with each theme. A summary of the associated number of topic ideas for each theme appears in Table 1 below. The apportioned number of associated topic ideas show that the themes reasonably represent the discussion of the meeting.

**Table 1: Topic ideas associated with themes**

| Theme | # Topic Ideas Associated |
|---|---|
| 1. Metadata, Standards, and Semantics | 15 |
| 2. Discovery, Search, and Recommendation | 9 |
| 3. Modernizing HDRL Users' experience | 11 |
| 4. Modernizing HDRL Staff's experience | 7 |
| 5. Modernized, Scalable, and Dependable Infrastructure for Compute and Storage | 10 |
| 6. Metrics and Impact Assessment | 6 |
| 7. Community Engagement and Sustainability | 8 |
| Total Topic Idea Associations | 66 |

* The total number of ideas was 53 but some topic ideas were associated with more than one theme for a total of 66 topic ideas associated with the themes.

[10] We presume the user meant IHDEA instead of HDRL since HDRL has no such recommendations.

We utilized the associated topic idea ratings to calculate ratings for each theme. The theme rating was calculated using the averaged value of associated topic idea ratings. An error estimate on each of these theme ratings was calculated by propagation of the estimated sampling errors of each topic idea rating using the survey sampling error[11]. We utilized the survey sampling error estimation using a population of 40 attendees and a 95% confidence level with the number of votes for each topic idea as the sample size.

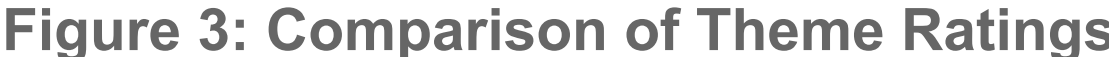

**Figure 3: Comparison of Theme Ratings**

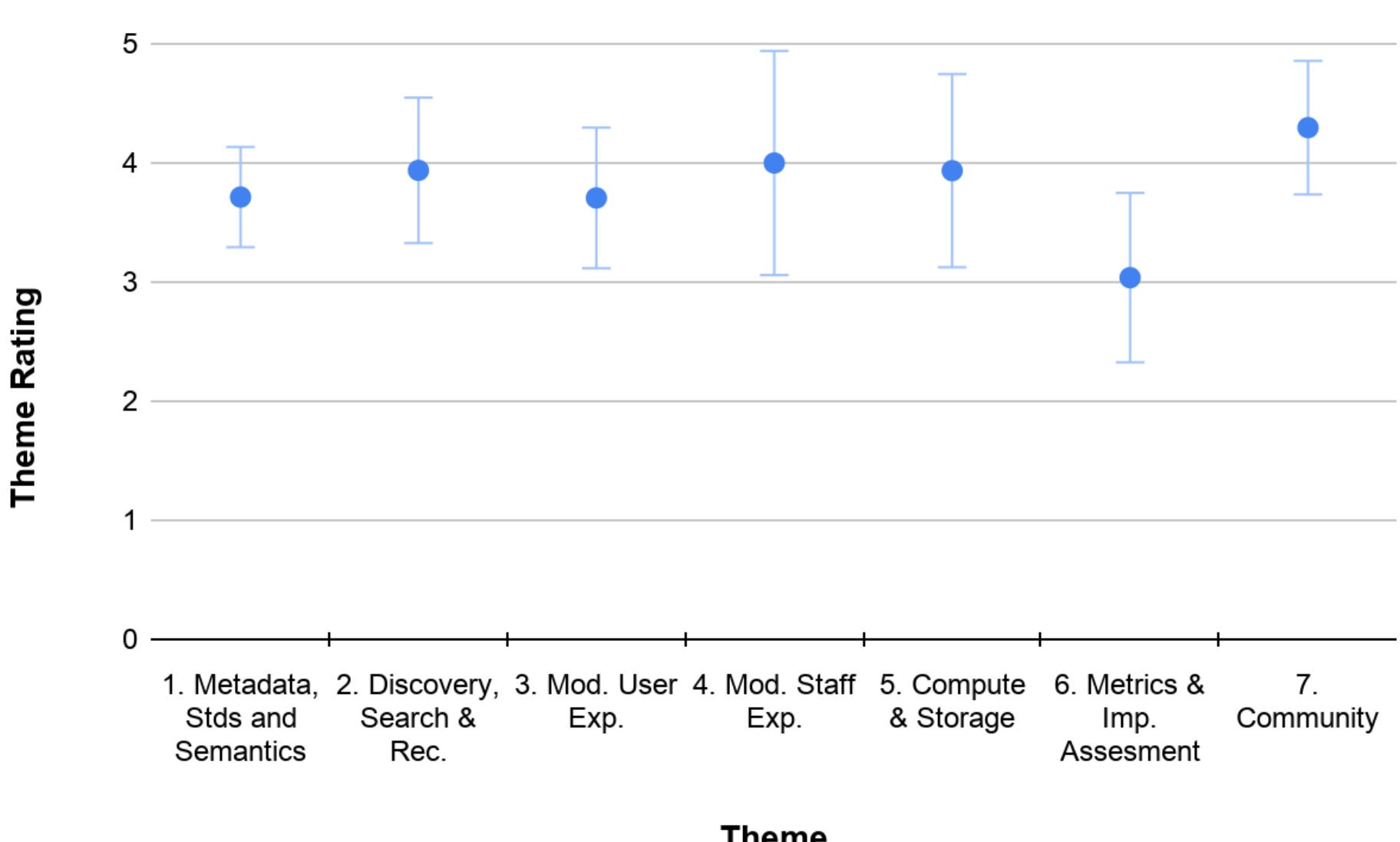

***Figure 3: Comparison of Theme Ratings.*** *All themes are equally rated by associated topic ideas. Values and error derivation are as noted in the text.*

Figure 3 shows a comparison of the derived theme ratings and 95% confidence intervals. This data appears to indicate that participants felt equally strongly about all of the themes and were interested in engaging in a broad discussion on issues impacting HDRL not just in, for example, the area(s) where they currently work.

The distribution of ratings in each theme was also examined (Figures 4, 5 and 6). Each histogram below shows the number of topic ideas in each theme with the given rating using rounded integer values. The average rating across all 53 topic ideas was 3.78 +/- 1.00, which would be shown as a 4 star rating in the distribution plots below. All plots show the number of ideas per star level for the indicated theme with identical vertical axis ranges.

---

[11] As calculated by the margin or error calculator on https://www.surveymonkey.com/. https://www.surveymonkey.com/learn/research-and-analysis/margin-of-error-calculator/

**Figure 4: Theme 1, 2, and 3 idea rating distributions**

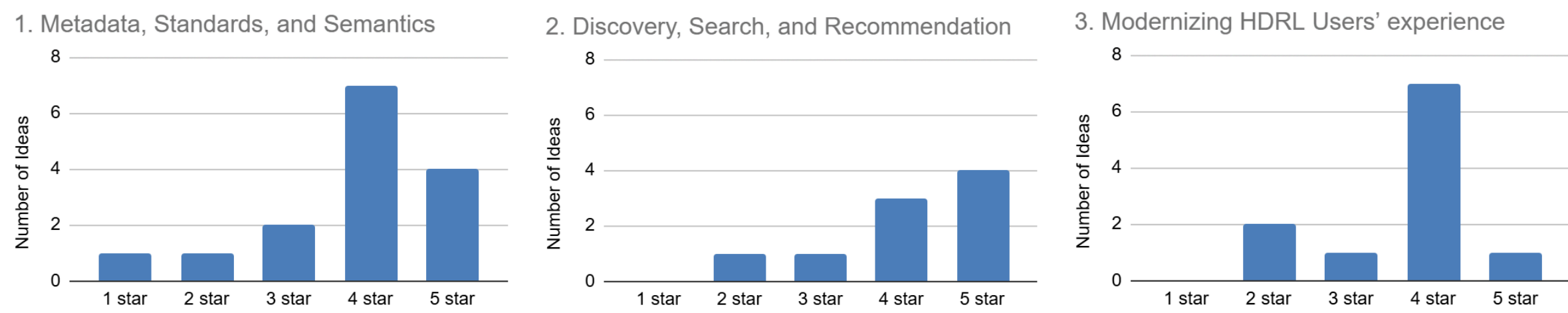

**Figure 5: Theme 4 and 5 idea rating distributions**

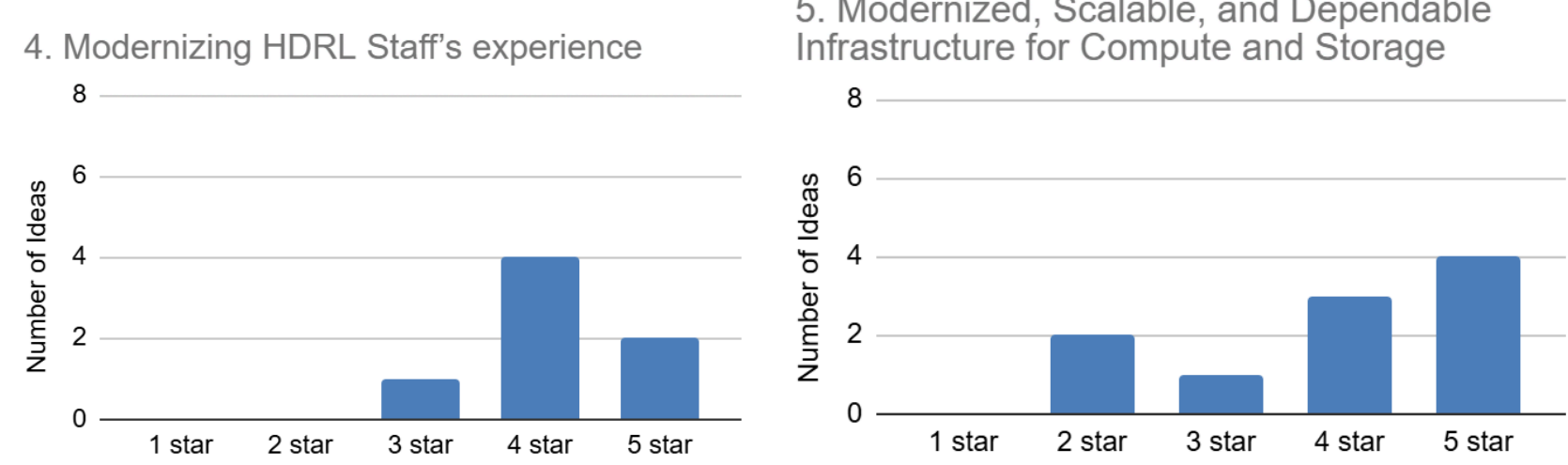

**Figure 6: Theme 6 and 7 idea rating distributions**

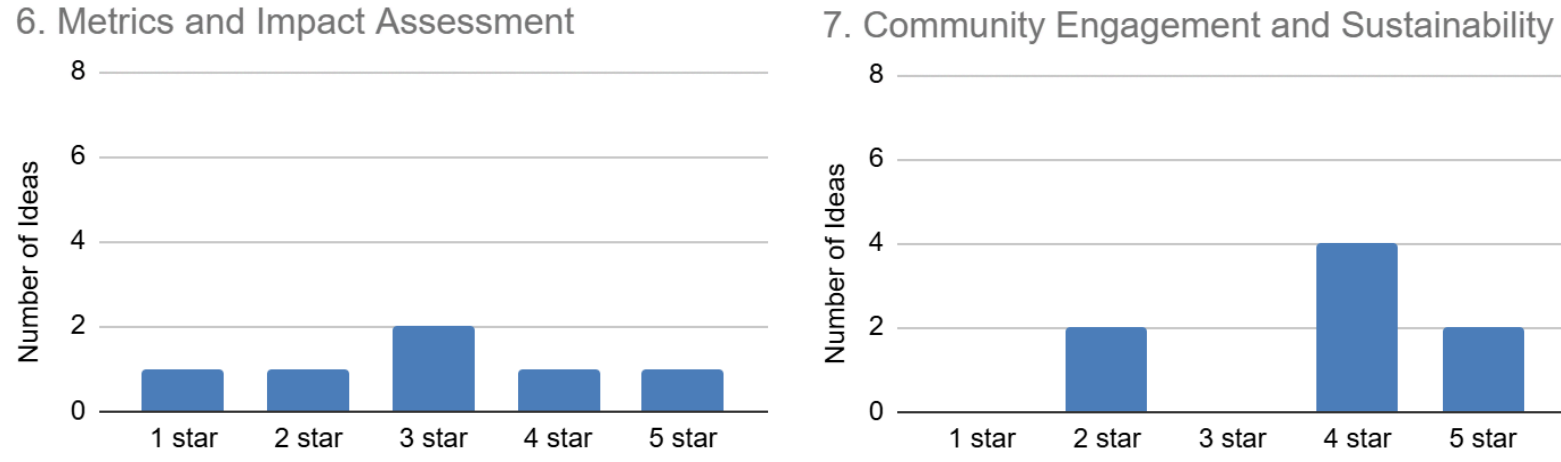

Cross comparing these distributions leads to a few insights. First, the distributions generally indicate a 'right-leaning' set of related topic ideas for all except theme 6 which has a broad and flat distribution. This may indicate the staff had a more or less cohesive enthusiasm for the topic ideas associated with most of the themes but had at best a mixed reaction to the topic ideas related to theme 6.

Finally, we examined those topic ideas which would fall into the "5-star" column in each distribution (Table 3). These 14 highly rated topic ideas provide some of the highest potential paths forward for each theme. Visually comparing the number of highly rated topic ideas for each theme (blue bars in Figure 7) and those actually discussed in the meeting (red bars) shows reasonable coverage for all themes except theme 1.

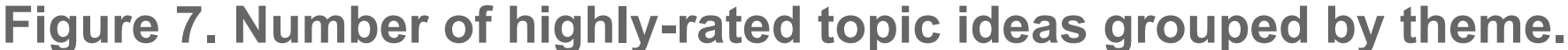

### Figure 7. Number of highly-rated topic ideas grouped by theme.

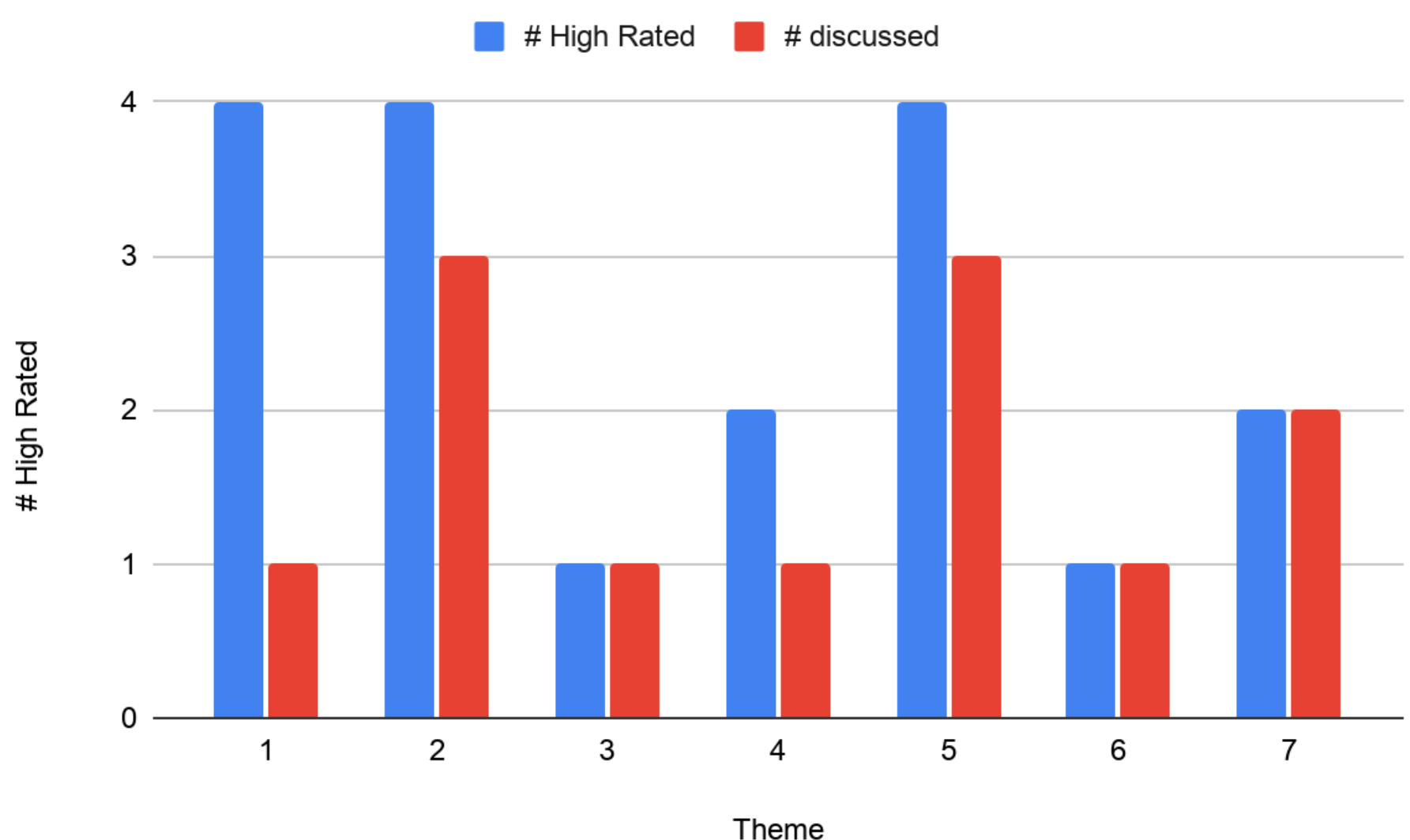

***Figure 7. Number of highly-rated topic ideas grouped by theme.*** *Topic ideas selected based on the topic idea rating of 4.5 or higher and then grouped by associated theme (blue bars). Red bars indicate the topic ideas rated 4.5 or higher which were actually discussed (see also table 2). Topic ideas associated with more than one theme are included in the data for both associated themes for a total of 18 overall.*

### Table 2: "5-Star" topic ideas in each theme

| Theme | Topic Idea ID |
|---|---|
| 1. Metadata, Standards, and Semantics | 5, **13**, 17, 54 |
| 2. Discovery, Search, and Recommendation | 17, **19, 36, 46** |
| 3. Modernizing HDRL Users' experience | **43** |
| 4. Modernizing HDRL Staff's experience | 5, **43** |
| 5. Modernized, Scalable, and Dependable Infrastructure for Compute and Storage | **18,** 27, **52, 55** |
| 6. Metrics and Impact Assessment | **36** |
| 7. Community Engagement and Sustainability | **9, 51** |

*Ideas discussed during the meeting are indicated in bold.

*Breakout Discussion Topics*

The breakout discussion content and associated statistical properties are shown in Table 3. In general, breakout discussions focused on individual topic ideas as indicated. The breakout discussion topic is given in the first column, followed by the topic idea ID, the average rating, the number of votes and the interest score for that idea in columns 2-5. The associated themes for that topic idea are given in the last column. Two breakout discussion topics covered multiple ideas, specifically the SPASE and Metrics topics, and the multiple topic ideas covered in those breakout discussions are listed in column 2 with topic ideas with a rating of 4.5 or greater indicated in bold text. In those cases, the rating and interest score were averaged across all the topic ideas specified, the number of votes were summed, and the list of associated themes were combined. Discussion topics involving more than one topic idea are indicated with an asterisk. For comparison, the average rating across the entire set of ideas is included on the penultimate row followed by the standard deviations in those values on the final row. Values higher than one standard deviation above the average are indicated in italicized text, and the maximum value for each column is emphasized with bold text. Topics that were discussed in two breakout groups are indicated with bold text.

**Table 3: Statistical Properties of the Highest Ranked Breakout Discussion Topics**

| **Breakout Discussion Topic** | **Topic Idea Property** | | | | |
|---|---|---|---|---|---|
| | **ID** | **Avg. Rating** | **# Votes** | **Interest Score** | **Associated Themes** |
| **CME Tracker** | **18** | **4.75** | ***12*** | ***57*** | 5 |
| **Formal Feedback** | **9** | 4.64 | *11* | *51* | 7 |
| **HDRL Workshops** | **51** | 4.55 | *11* | *50* | 7 |
| HDRL Data Knowledge Graph | **36** | 4.67 | *9* | *42* | 2, 6 |
| SPASE* | 4, **43**, 30, 33, 32, 6 | 4.11 | 47* | 32.5 | 1, 3, 4 |
| Drafting SPASE records from Docs | 33 | 4.17 | 6 | 25 | 3, 4 |
| Serving Big Data | 50 | 4.20 | 5 | 21 | 5 |
| Metrics* | 35, 34, 44 | 3.12 | 18* | 21 | 6, 5 |
| **Avg Value for All Ideas** | | **3.78** | **6.11** | **24.08** | **—** |
| **Standard Deviation** | | **1.00** | **2.76** | **13.41** | **—** |

* The rating and interest score were averaged across all the topic ideas specified, while the number of votes were summed.

Table 3 above shows the interest of the attendees in each topic as evidenced by the selection of the topic ideas for discussion in one or more breakouts and by the associated voting statistics. While higher values of the calculated interest score and average rating appear correlated with what was selected by the attendees, it is not entirely deterministic of what was selected.

Furthermore, we note that the calculated survey margins of error for topic ideas are not useful for statistically distinguishing between them due to low sampling statistics. The highest voted topic ideas had a margin of error of ~25% (12 votes), with much higher margins of error for the lower voted topic ideas (up to a 98% margin of error!). Only those topic ideas with a high number of votes and very different average ratings can be statistically distinguished from one another. In many cases we simply can't make statistically meaningful comparisons in the ratings between topic ideas due to the low number of votes.

*Mapping Breakout Discussions to Themes*

The topics that were discussed in more than one breakout group (bold-faced in column 1 of the above table) were assigned to theme 7 ("Community Engagement and Sustainability") and theme 5 ("Modernized, Scalable, and Dependable Infrastructure for Compute and Storage"), providing examples of well-supported projects in those areas. Two discussion topics with comparably slightly enhanced statistics cover the remaining five themes, namely the "HDRL Data Knowledge Graph" and "SPASE", providing potential paths forward in support of those themes. The next two topics in Table 2 directly support the four already mentioned. The "Metrics" discussion topic, from the staff perspective, was not well supported.

## Discussion

### High Level Results

The gathered meeting metrics can help to organize and prioritize the large amount of generated content. The 53 topic ideas submitted by workshop attendees are reasonably categorized into seven themes with fair representation in each as shown by the distributed associated topic ideas for each theme (Table 1). These themes are all relevant for the HDRL mission to acquire, curate and disseminate NASA heliophysics data (Roberts et al. 2023).

We find the participants felt strongly supportive of all the themes (Figure 3) and the generated topic ideas indicated that attendees were interested in discussion of a broad range of relevant subjects and not just those related to their area of work. There were many highly rated topic ideas generated for each theme with the exception of theme 6, Metrics and Impact Assessment. This may indicate lower attendee enthusiasm in this area and the need for further followup to better define what associated ideas or projects in this area would be of value (as seen in Figure 4).

It is challenging to distill down this content to find what the attendees felt is the most impactful direction(s) for HDRL to consider. During the meeting, topic ideas were sorted by their "interest score” (see Tables 4 and B1), but this was not always the best metric to determine what would actually be discussed (Table 3). Further, the differences between similarly rated ideas were statistically indistinguishable from one another.

Selection of breakout topic ideas is an obvious approach to identifying impactful directions, including the tailoring of the discussions by session attendees. For example, two breakout sessions combining several associated topic ideas of varying average ratings into a single discussion (SPASE and Metrics breakout discussions), and some topic ideas were selected more than once (see Table 3). Both cases indicate enthusiasm by attendees in these areas.

However, selection of breakout topic ideas is also an imperfect metric. For various reasons, some highly-rated topic ideas (notably those in theme 1, see Figure 7) were not picked up for breakout discussion while others were selected more than once. Looking at this information, we conclude that the selection guidance needs improvement and the interest score is less valuable than anticipated. The topic idea average rating with some minimum number of votes appears to be a better metric for measuring staff support for topic ideas.

Nevertheless we may consider each of the factors together to help provide a list of prioritized ideas for followup (Table 4). We constructed this list of ideas to include well-supported topic ideas, topic ideas selected for breakout discussion, and topic ideas considered directly relevant for existing challenges. The associated topic ideas in column 2 include those directly related to the stated idea but also include topic ideas considered necessary to support, enable, or accelerate the stated idea, and are followed by the list of themes associated with those topic ideas in column 3. All seven associated themes are covered by the chosen ideas.

Some of these ideas are complex in structure, such as “Formal Feedback” and “Build a Metrics System”, or are representative of more general efforts needed, such as “CME Tracker”.

The “CME Tracker”, “Formal Feedback”, and “HDRL Workshops” topics were all selected for discussion twice, indicating broad interest in those ideas and their wide scope. Perhaps unsurprisingly, the discussions of these three topics were interwoven as suggested in Figure 7.

Based on those discussions, one possible way to address all three is to create a dedicated HDRL user workshop where attendees’ travel expenses are paid in exchange for detailed user feedback on the HDRL services they use to analyze a CME or other phenomenon occurring in HDRL-hosted data. Such a workshop would ideally be supported by trained usability testing personnel to improve feedback quality. Another flavor of this workshop described in the discussion summary would be a tutorial of a relevant HDRL service at a community meeting (e.g., a mission-specific workshop) followed by hands-on mentoring with user feedback. More details on the selected topics are available in the sections summarizing the associated breakout sessions and the associated topic ideas in Appendix B (see Table B1).

**Table 4: Prioritized Ideas**

| Idea Title | Associated Topic Idea IDs | Associated Themes |
|---|---|---|
| CME Tracker (“Find the Gap in HDRL”) | 18 | 5 |
| Formal Feedback | 9, 22 | 7 |
| HDRL Workshops (“Bring HDRL to the People”) | 51 | 7 |
| HDRL Data Knowledge Graph with Supporting Semantic Structures | 13, 36 | 1, 2, 6 |
| SPASELite Editor | 30 | 3 |
| Build a Metrics System for HDRL | 34, 35, 40 | 5, 6 |
| A Hybrid HPC/HelioCloud for Big Data Support | 50, 52 | 5 |
| Mission Data Delivery to HDRL | 27, 48 | 3, 5 |
| Using AI Approaches to Accelerate Curation and Discovery, including SPASE* | 6, 19, 32, 37, 42, 43, 46 | 2, 3, 4 |
| Improve Reusability and Trust of HDRL Data for New Users | 17, 25, 26, 55 | 1, 2, 5 |
| Streamline Event Data Support in the HEK | 5, 54 | 1, 4 |

*Renamed for clarity

These topic ideas may also be interdependent upon each other. Examples of those connections are shown in Figure 8 with ideas considered to be primarily infrastructural colored green with others indicated in grey. Arrows indicate what other selected ideas would likely benefit from work in that area. For instance, developing Heliophysics semantic resources and streamlining their use with approaches such as keyword vectorization is expected to improve our support for event data, SPASE, and the HDRL data knowledge graph by connecting the metadata in each through a unification of terminologies (e.g., phenomena, classes of data).

Investing resourcing in HDRL’s cloud and hybrid strategy is expected to provide additional infrastructure to support our service integrations (“Find the Gap in HDRL”), streamline mission data delivery to HDRL, and increase our support for HDRL workshops. Connections between these ideas and other HDRL infrastructure and services are not shown but are important. For example, work on the CME tracker idea is intended to uncover the gaps in services researchers use at HDRL to perform that research resulting in assigned work to improve those services.

**Figure 8: Example connections between the prioritized ideas**

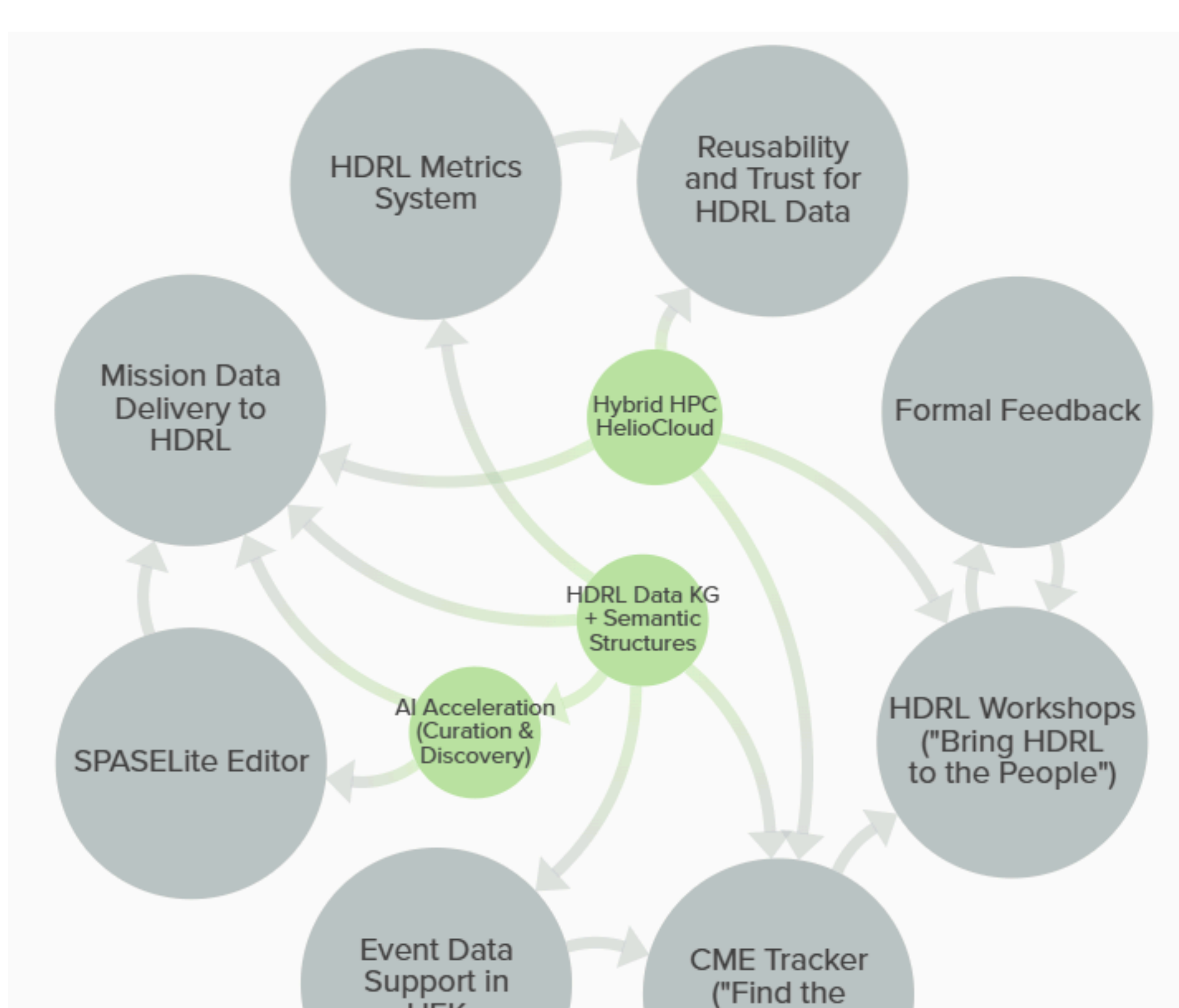

***Figure 8: Example connections between the prioritized ideas****. Ideas are shown with grey and green circles. Those considered infrastructural are shown in smaller green circles. Arrows indicate what other items are expected to benefit from work in that area.*

The interconnections help clarify that these prioritized topic ideas are in fact aspects of broader areas where it would be productive to take action. Considering the gathered information we conclude the following broader strategic prioritized areas exist:

- Improve the mission data delivery process to HDRL;
- Formulate and incorporate formal feedback methods into relevant interactions with HDRL stakeholders, both automated and manual;
- Develop and implement a balanced outreach and engagement plan, including HDRL Workshops;
- Investigate means to better capture useful metrics related to the HDRL mission;
- Streamline and modernize HDRL's operations and services, incorporating AI/ML approaches where beneficial;
- Prepare and develop solutions for serving big data, including cloud and HEC solutions;

- Engineer a semantic framework for Heliophysics resource interlinking, working towards a HDRL data knowledge graph;
- and
- Address gaps in Heliophysics data environment standards, in particular with heliophysics events and in semantics.

## Lessons Learned

Overall, the feedback from attendees concerning the meeting was positive. They valued the networking opportunity with other HDRL staff and grantees, the high-level and quick-paced talks covering a wide range of services and efforts at HDRL, and the interactive discussion sessions based on the Padlet ideas. Attendees that provided feedback agreed on repeating the meeting on a regular cadence, such as every 2-3 years, to stay informed about HDRL's work and to connect with colleagues in person.

Attendees noted the breakout sessions spent too much time selecting the discussion topics and an overall lack of organization in those sessions. In addition to it taking significant time, the selection process was somewhat fraught. For example, nine topic ideas with lower ratings were selected for discussion instead of other topic ideas with higher average ratings or more votes, indicating a potentially less efficient use of discussion time than planned (see column 2 of Table 3). This may be due in part to topic ideas being selected for discussion based on preference by the more outspoken members of the breakout group.

The selection of lower rated topic ideas for discussion could also point to the statistical similarity of the topics due to the high margin of error on topic idea ratings (at least 25% margin of error for highest voted topic ideas and much worse for other topic ideas with fewer votes). Only topic ideas which were most highly rated could be distinguished from the lowest rated ideas. More votes per topic idea are needed to decrease the margin of error for each and thus more meaningfully distinguish between the ideas during discussion topic selection. For instance, an average ~15% margin of error on a 1-5 point scale would give the ability to statistically distinguish between 'high', 'medium' and 'low' ratings.

Additionally, several highly rated topic ideas were not discussed in the meeting (theme 1 in Figure 7) showing that additional discussion time could also be more useful in another iteration of the meeting.

Several suggestions were given in attendee feedback to improve the logistics and organization of various portions of the meeting. More structure was requested for the lightning talks, such as a presentation template, and a more streamlined approach for the recaps from the discussion breakout sessions, including asking the person who submitted the selected topic idea to explain the idea before the breakouts begin to improve clarity.

The use of the Padlet was well regarded, but attendees requested (1) that the ideas not be shuffled as they are entered to make it easier to follow, (2) instructions on how to vote with a "0" star rating, and (3) to have a "why" part of the topic idea, such as what existing challenge it could solve, to help others better understand its perceived importance. Additional flexibility in the schedule was also requested.

To address the feedback requesting more efficient use of the discussion breakouts, the attendee that submitted a topic idea selected for discussion could explain the idea before the groups separate into the breakout sessions. Then, this topic would be assigned to a given breakout session, ideally resulting in relevant staff joining that group. The requested additional flexibility in the meeting agenda could be supported by shifting the recaps from the discussion sessions to be after a break and by adding a "wild card" session at the end of the meeting determined by popular vote. Topics favored by the attendees would then be discussed in that breakout discussion in the same format as the others.

## Summary & Next Steps

All three of the primary goals of the meeting were successfully met. Staff awareness of HDRL and related activities was achieved via the presentations and discussions which followed which were demonstrably broadly engaged in by participants. Meeting attendees were also (re-)exposed to the drivers of HDRL strategy and participated in several exercises to consider gaps in the strategy and potential solutions. The copious output of the meeting attendees and positive meeting feedback also indicates significant engagement and success with the final goal to leverage HDRL staff and grantee expertise to better understand HDRL strategic gaps and solutions.

Seven themes were identified in a subsequent analysis of the attendee feedback as equally well-supported by the meeting attendees:

1. Metadata, Standards, and Semantics
2. Discovery, Search, and Recommendation
3. Modernizing HDRL Users' experience
4. Modernizing HDRL Staff's experience
5. Modernized, Scalable, and Dependable Infrastructure for Compute and Storage
6. Metrics and Impact Assessment
7. Community Engagement and Sustainability

The plethora of topic ideas associated with these themes were reduced to 11 prioritized ideas (Table 4) based on the associated topic's average rating, its number of votes, whether the topic was discussed during the meeting, and whether the topic was considered directly relevant for known challenges or drivers.

Considering the gathered information and the interconnections between the prioritized ideas we conclude the following broader strategic areas for action exist:

- Improve the mission data delivery process to HDRL;
- Formulate and incorporate formal feedback methods into relevant interactions with HDRL stakeholders, both automated and manual;
- Develop and implement a balanced outreach and engagement plan, including HDRL Workshops;
- Investigate means to better capture useful metrics related to the HDRL mission;
- Streamline and modernize HDRL's operations and services, incorporating AI/ML approaches where beneficial;
- Prepare and develop solutions for serving big data, including cloud and HEC solutions;
- Engineer a semantic framework for Heliophysics resource interlinking, working towards a HDRL data knowledge graph; and
- Address gaps in Heliophysics data environment standards through IHDEA, the international standards body for Heliophysics, in particular for heliophysics events and semantic resources.

Overall, these ideas are considered strongly supported by HDRL staff interest, relevant to existing challenges, and are expected to positively impact HDRL's stakeholders, including HDRL staff and the Heliophysics research community.

As of this writing, some work has already begun for some of these ideas.

1. Discussions are now in progress with the NCCS to begin work on hosting a HelioCloud instance on HEC resources there.

2. An AI tool developed to determine usage of data collected by SOHO and SDO in peer-reviewed publications is now being implemented which may improve reusability and trust in those data and provide useful metrics.

3. HDMC staff are also preparing to develop AI approaches to extract SPASE metadata from text and data files based on existing work, which is expected to accelerate the creation and curation of SPASE records.

4. New IHDEA working groups around heliophysics semantics and event standards have been formed with HDRL participation. An IHDEA working group for definition of a standards process in IHDEA, co-led by HDRL, has also been initiated with the aim to provide concrete processes to advance these efforts.

5. Finally, HDMC held a half-day debriefing on the following Friday (see Appendix D) to further capture thoughts from the meeting.

It is clear that more actions may be taken based on these results. In addition to further evaluation of the results of this meeting we expect further followup via future focused mini-workshops or other engagement in order to better understand and flesh out some of the more complex and interdependent ideas.

The feedback by attendees indicated that they felt the HDRL Strategy meeting to be a valuable use of their time (see Figure C1). By attendee request, the next HDRL Strategy meeting is expected in 2028 or 2029, possibly extended by a half-day to provide more discussion time and flexibility in the schedule as requested. Several improvements are planned for the next HDRL Strategy meeting based on the challenges encountered in the analysis presented above and the feedback provided by attendees.

## Appendix A. Meeting Agenda

### Table A1. Meeting Agenda

| Day 1 - Jan 21 (Wed) | |
|---|---|
| **Session Description** | **Time** |
| Opening session<br>● Brian Thomas: Why are we here?<br>● Lan Jian: SPDF<br>● Jack Ireland: SDAC<br>● Brian Thomas: HDMC, HDRL<br>● Ice breaker | 9:00 - 9:20<br>**RECORDED SESSION**<br>Room 183A |
| Show and Tell: Current Infrastructure<br>● Tami Kovalick: SPDF data processes<br>● Ritika Prasai: SDAC data quality<br>● Bernie Harris: CDAWeb and SSCWeb<br>● Rita Johnson: SSC spacecraft orbit query capabilities<br>● *5 min question break*<br>● Raissa Woodland: Spacecraft 4D orbit viewer<br>● Daniel Briseno: Helioviewer<br>● Ed Mansky: Virtual Solar Observatory (VSO)<br>● Alisdair Davey: VSO 2.0 and ASDF<br>● *5 min question break*<br>● Mike Liu: CDF development<br>● Ryan Timmons: Heliophysics Event Knowledgebase<br>● Jeff Bradford: HSDCloud operations and infrastructure<br>● Sandy Antunes: HelioCloud development and the CloudCatalog<br>● *5 min question break* | 9:20 - 10:40<br>**RECORDED SESSION**<br>Room 183A |
| break | 10:40 - 11:00 |
| Show and Tell: Community Connections and Operations<br>● Rebecca Ringuette: Open Science at HDRL<br>● Jon Vandegriff: HAPI, Cloud Compute for CGS DRIVE center and PSP<br>● Wale Jaiyeola: HelioData<br>● Julie Barnum: Heliophysics Software Search Interface<br>● *5 min question break*<br>● Alex Young: Science Outreach<br>● Rose Brunning: Communications<br>● Raphael Attie: HelioNauts<br>● Chris Bard: HelioAnalytics<br>● *5 min question break* | 11:00 - 12:00<br>**RECORDED SESSION**<br>Room 183A |
| lunch | 12:00 - 1:30 |
| Show and Tell: Metadata and Data Management<br>● Andrew Robbertz: SolarNet<br>● Andriy Koval: ISTP and SPASE 3.0<br>● Eric Grimes: ISTP/SPASE metadata editor on HelioCloud and Plot Walk<br>● Bob Weigel: Coordinate systems, units, and mapping from ISTP to SPASE<br>● *5 min question break*<br>● Jen Sun and Bernie Harris: Powering the Heliophysics Data Portal with the SPASE database<br>● Tressa Kasulke: SMWT and SPASE Registration/Curation Efforts<br>● Zach Boquet: SPASE-DataCite and SPASE-soso softwares<br>● *5 min question break*<br>● Ryan McGranaghan: Improving connections through semantics<br>● Anthony Buonomo: AI-powered curation<br>● Aidan Scharnikow: Using the Model Context Protocol with science literature.<br>● *5 min question break* | 1:30 - 2:30<br>**RECORDED SESSION**<br>Room 183A |
| Break | 2:30 - 3:00 |
| HDRL Drivers *and Breakout Topics*<br>● Archive Research<br>● Ground and Space Data<br>● HelioSystems<br>● Gold Standard / Open Science<br>● Community Support | 3:00 - 3:20<br>Room 183A |
| Challenge Breakout<br>**Moderators**: Brian Thomas, Lan Jian, Jack Ireland, Rebecca Ringuette, Alex Young | 3:20 - 4:15<br>Room 183A, 183B, Wow room, offices |
| Report Out | 4:15 - 4:45<br>Room 183A |
| *Wrap-up* | 4:30 - 5:00<br>Room 183A |
| HDRL Social | 6:00 - till |

| Day 2 - Jan 22 (Thurs) | |
|---|---|
| **Item** | **Time** |
| *Voting on Padlet* (Challenges) | 9:00 - 9:45<br>Room 183A |
| break | 9:45 - 10:20 |
| *Breakout 1*<br>**Moderators**: Brian Thomas, Lan Jian, Jack Ireland, Rebecca Ringuette, Alex Young | 10:20 - 11:00<br>Room 183A, 183B, WOW room, offices |
| *Report out 1*<br>Short 5 min break | 11:00 - 11:25<br>Room 183A |
| *Breakout 2*<br>**Moderators**: Brian Thomas, Lan Jian, Jack Ireland, Rebecca Ringuette, Alex Young | 11:30 - 12:00<br>Room 183A, 183B, WOW room, offices |
| *Report out 2* | 12:00 - 12:30<br>Room 183A |
| lunch | 12:30 - 2:00 |
| *Reconvene*<br>Choose what topics to discuss next. | 2:00 - 2:15<br>Room 183A |
| *Breakout 3*<br>**Moderators**: Brian Thomas, Lan Jian, Jack Ireland, Rebecca Ringuette, Alex Young | 2:15 - 3:00<br>Room 183A, 183B, WOW room, offices |
| *Report out 3* | 3:00 - 3:30<br>Room 183A |
| break | 3:30 - 3:45 |
| *Reconvene/Plenary*<br>Full group discussion on the meeting content, and suggested priorities | 3:45 - 5:00<br>Room 183A |

## Appendix B: Topic Ideas and Voting Results

This appendix contains all of the relevant information concerning the generated topic ideas for the meeting. Associated emergent groups, identified in the Analysis section, are also shown mapped to these topic ideas. Rows shown with boldfaced text were discussed in the meeting.

### Table B1. Meeting Topic Ideas

| Idea ID | Title | Description | Assoc. Themes | Avg. rating | # of Ratings | Interest score |
|---|---|---|---|---|---|---|
| **18** | **Put together a CME tracker** | **Look at what the St. Patrick's day storm papers used for data, software and services. Similar for the Gannon storm in May 2024. Build a prototype to figure out where the infrastructure gaps are and fix them.** | **5** | **4.75** | **12** | **57** |
| **9** | **Formal Feedback** | **Need a common place to gather and save feedback from different people and events. Possibly create a group / meeting to go through different feedback. Possibly bi-monthly. Make it more formal to collect feedback. Shared document or MIRO board.** | **7** | **4.64** | **11** | **51** |
| **51** | **HDRL Workshops** | **Japan (Iugonet) regularly hosts data analysis workshops for people to learn how to analyze the data they serve. Should we have an HDRL flavor of the PyHC workshop, where all the tutorials involve data served by HDRL, preferably with other data?** | **7** | **4.55** | **11** | **50** |
| 17 | How to trust/use new-to-you data? | Trusting data from an instr you've never worked with. Include links to review papers (more vetted than just any paper)? Code gen via previous examples? | 2, 1 | 4.50 | 10 | 45 |
| 4 | Discuss what needs to be done to finish ISTP/ HAPI/ SPASE mapping | *No data* | 1 | 4.40 | 10 | 44 |
| **36** | **Build a knowledge graph of HDRL data** | **Building a knowledge graph of the connections between HDRL datasets helps us (and HQ) understand what data are important or foundational to other data. This should include connections across missions through publications where multiple datasets are used. We can lean on the Sci-X services for this. Doing this helps us create informed impact metrics for our data, which can then be reported to HQ.** | **2, 6** | **4.67** | **9** | **42** |
| 43 | Use vectorization for keywords | Use vectorization to determine similarities between keywords across multiple vocabularies, including GCMD, UAT, AGU Index. The intended application is for this tool to be used to then suggest keywords for SPASE records from those vocabularies based on the given text (e.g., description, abstract, documentation, etc). Those keywords can then be reviewed by the data producer / HDRL curator for accuracy. | 4, 3 | 4.56 | 9 | 41 |
| 30 | Create a SPASELite editor for NASA | Create a SPASE editor with a built-in validator to be used by researchers or scientists supported by other US agencies to create SPASE records fulfilling the minimum NASA SPASE requirements, possibly with ~3 other easy | 3 | 4.00 | 10 | 40 |

| | | fields we consider useful on HelioData for search. This new editor MUST be easy to use and intuitive. | | | | |
|---|---|---|---|---|---|---|
| 11 | Create a Community Resource Page | Build an HDRL site “featured tools” area. Sort HDRL resources by area (software, search, archive, etc) with one line “how can this help you” and a link directly to the site.<br>APIs and a list of information and a one line description of what it does for you. Incorporate multiple partners and institutions. | 3 | 4.00 | 9 | 36 |
| 15 | Look to earthsci | How does earth science handle multiple instr types & sources so researchers can discover and use it? | 2 | 3.50 | 10 | 35 |
| 23 | Funding Issues | It's still the same problem we have always had and was hinted at by a few talks this morning. You can get funding for really cool ideas that create useful tools, but getting money to keep them going is hard. How do we keep these projects going? Or, how do we get funding panels to not penalize continuation proposals just because they are not necessarily proposing new and shiny things? | 7 | 4.38 | 8 | 35 |
| 26 | LLM over services | Want to add natural language queries on top of our services, including the magnetic conjunction notebooks, the 4D orbit viewer, HelioViewer, CDAWeb, etc. | 2, 5 | 4.25 | 8 | 34 |
| **35** | **Build a metrics system for our repositories** | **Also need a real-time monitoring system for these types of metrics (e.g., data volume, number of users, number of accesses, etc) for HDRL as a whole and SPDF and SDAC separately. Others said this decreased the number of requests from HQ with quick turn-arounds. Need to include all websites in these metrics (SDAC, SPDF, helio.data, etc).** | **5, 6** | **4.25** | **8** | **34** |
| 42 | Strategizing for discovery by AI | Need to understand what metadata AI tools (e.g. Gemini) use to recommend datasets to users. Is it embedded schema.org metadata, is it related to search engine optimization, does Croissant metadata improve this? Answering these questions will help prioritize what metadata we need and where it needs to go. Bonus is to figure out how we can feed those tools information about the APIs and dataset documentation to build example analyses for our data. | 2 | 3.78 | 9 | 34 |
| 13 | Create classes of data? | Define classes of data then tag datasets, so scientists can find datasets new to them. | 1 | 4.71 | 7 | 33 |
| 19 | Recommendation engines | Eg I look up a specific Stereo CME because I know STEREO, it would be nice if the search also links to other missions that saw it. (VSO2 exploring this?) | 2 | 4.57 | 7 | 32 |
| 29 | HDRL[12] recommendations | We need to create a way to propose an HDRL recommendation. At the 2025 IHDEA meeting, we discussed using the IVOA as a template and possibly simplifying it. The two pending recommendations are for units and coordinate frames. | 7 | 4.00 | 8 | 32 |
| 38 | Determine useful DataCite metadata richness | Work with Sci-X to determine what DataCite fields they use in their system, what additional fields we might find useful, and prioritize what fields we (HDRL) should focus on populating. Then, we can compare this to our currently implemented mapping to DataCite and fill in any holes. We can also compare this to our current metadata quality plans to prioritize what is needed (beyond the basic citation fields). | 1 | 3.88 | 8 | 31 |

[12] It is likely that instances of “HDRL” should be read as IHDEA for this topic idea. HDRL is a member of IHDEA and adheres to IHDEA standards. The creation of a process for IHDEA standards has been initiated based on this topic idea.

| | | | | | | |
|---|---|---|---|---|---|---|
| 52 | What would a move into HPC look like? | Serving data on the cloud up to 1.5 PB seems to be workable, but we will soon be hosting larger datasets that researchers will want to perform initial analyses on. How do we support this? Would this look like installing HelioCloud on a NASA HPC for serving large data and colocated analysis? Would it be useful to incorporate HySDS to link to the cloud? | 5 | 5.00 | 6 | 30 |
| 48 | "Submit" button on HelioData | To simplify the artifact submission process in Heliophysics, we could create a 'submit' button on the Helio.data website. The button would have a few questions that will help us determine where the artifact should be submitted to, and then the request would be sent to the right repository (e.g., SDAC, SPDF, or our official Zenodo community) and the correct submission instructions could be sent to the requestor. | 3 | 4.14 | 7 | 29 |
| 33 | Draft SPASE records from SDAC/SPDF metadata using AI | Use an AI tool to draft a mapping from SPDF and SDAC metadata systems into SPASE. Once this mapping is reviewed, an AI tool can be used to draft an implementation of that mapping in Python. Can then use that tool to draft SPASE records from those metadata systems. | 1 | 4.33 | 6 | 26 |
| 14 | Warning labels for data | Instr-specific scientist friendly metadata, eg 'SOHO particle data is taken at L1' or 'Wind particle data is taken while spinning'. VSO might add popups. Also always show metadata. | 1 | 3.57 | 7 | 25 |
| 25 | Adding data available through TAP | What will it take to incorporate TAP into HelioViewer to show data from Europe in the same services as FITS files? | 3 | 4.17 | 6 | 25 |
| 27 | Better file flow into HDRL | Require missions to push changes to HDRL to avoid us missing updates. Better file validation is also needed so the files port seamlessly into our systems. | 5 | 5.00 | 5 | 25 |
| **32** | **Draft SPASE records from documents using AI** | **Partner with other divisions, including Earth Science, to create an AI tool to extract metadata for datasets from documentation, related publications, OSDMPs, etc. The helio-specific need would be converting the output into a valid draft SPASE record.** | **3, 4** | **4.17** | **6** | **25** |
| 7 | Save dmsp data | *No data* | 7 | 4.00 | 6 | 24 |
| **34** | **Build a metrics system for datasets** | **Collaborate with Sci-X to build a metrics system for HDRL data, to be incorporated on the landing pages on helio.data. Will also likely include components from Google Analytics or similar for metrics about downloads, accesses, etc.** | **6** | **3.43** | **7** | **24** |
| 20 | Service maturity levels | Some indicator of how mature / trustworthy a given service is so a researcher knows what is worth spending their time on using. Something like PyHC levels (gold, silver, bronze) as part of a standard system for services (eg documentation, citation). Uptime / stability too. | 5, 3 | 3.14 | 7 | 22 |
| 3 | Develop an official connection between missions and the HDRL software / tools community | Work within SMD and Heliophysics specifically to incorporate call for community tool development into mission calls. At least get mission calls to include language about leveraging community tools and engaging existing HDRL elements. | 7 | 4.20 | 5 | 21 |
| 10 | Make it easier to find data | Make metadata leverage schema.org | 1 | 3.50 | 6 | 21 |
| **50** | **How do we prepare to serve big data?** | **HAPI silently fails when serving big data. Do our services have the same problem? When will a service like Globus be needed? Do we need to shift to HPC for the largest datasets coming? Need to look into** | **5** | **4.20** | **5** | **21** |

| | | **this.** | | | | |
|---|---|---|---|---|---|---|
| 39 | Determine SEO strategy | Work across divisions to determine a useful Search Engine Optimization (SEO) strategy for our dataset landing pages based on their experiences. This will likely include structured data (e.g., Schema.org JSON-LD), keywords on the page, and so on. Once this is determined, the metadata fields prioritized in SEO can also be prioritized in our metadata quality work. | 1, 7 | 2.22 | 9 | 20 |
| 6 | SPASE Statistics Chat Bot | Related to other entries listed here but would like a user friendly interface which allows querying about certain SPASE metadata fields of interest. Can use the HelioData database/api to help with this. | 4 | 3.17 | 6 | 19 |
| 5 | Rapid first ingest of event lists in HEK for standardization | Adding new event sources into the HEK can get bogged down in discussions with the provider about how their metadata maps to our schema. More tool improvements from us could help, but would there be value in allowing a first pass through an event list with a bare-bones set of attributes to get it visible/searchable first, then return later to enrich each entry? This would help with the “Helioviewer holy grail” idea as Helioviewer is using HEK API as one event source. | 1, 4 | 4.50 | 4 | 18 |
| 8 | Make Helioviewer the Holy Grail | If your data is not available in Helioviewer, it's worth investing resources in making it available in Helioviewer. Today on Helioviewer, you can browse to a specific date, view events from the HEK, download the science from VSO, find papers about an event via ADS, create a timelapse animation for a specific event using images from several different observatories, and share a link to Helioviewer to help someone else browse.<br><br>If there has been a WSA-ENLIL model run for a CME, you can click the event and view how that CME propagates through space and which spacecraft or planets are impacted thanks to CCMC's APIs.<br><br>During this meeting I see that we have so much more information that can be presented here.<br><br>Imagine: Looking at the same event, you can use the Data -> Paper lookup to find papers about this specific event, or related phenomena. Then we extend to our Paper -> Data infrastructure to find related datasets that may be of interest, and it cites "Provided by HDRL AI"<br><br>Maybe when browsing earth based data like SDO, you can see pointers to where STEREO is and where Solar Orbiter is thanks to SPDF's SSCWeb which already powers the 4D orbit viewer, then you know oh, we have observations for the backside of the sun for this time. When hovering over these pointers it shows a link stating "Provided by SPDF's SSCWeb"<br><br>While you're looking at any given observation, there can be a recommendation button which shows what other observations you may be interested thanks to the ongoing curation efforts with linking related datasets and it states "Provided by SPASE/HDRL Community"<br><br>Helioviewer can do a lot, but it can't do everything. So when we link back to data, we can state "Check out the | 2 | 2.83 | 6 | 17 |
| | | Heliophysics Software Search site for tools to help you analyze this data"<br><br>If we bring these services into Helioviewer, then from there everyone could get a good idea of what data and services are available. | | | | |
| 37 | Create a quality assessment tool using AI | Create a quality assessment tool to judge quality for text based fields, particularly the title and description fields. We have plenty of high quality examples to train it on. | 4, 3 | 4.25 | 4 | 17 |
| 22 | Need some natural language to report service problems | Need an easier way for people to report problems with services, maybe the analysis working groups or a chatbot? | 3 | 4.00 | 4 | 16 |
| 40 | Develop our own Open Science compliance report for publications | The numbers from Steve Crawford's presentation on Open Science compliance in NASA HPD publications are not favorable. A more considered approach for Helio using the same compliance criteria will likely improve those metrics. Can use the Sci-X services for some, but may need to contribute to some improvements in their algorithms. Can then include this on a metrics page in real time hosted somewhere (the hdrl website?). | 6 | 3.20 | 5 | 16 |
| 54 | Create a standard way to access event lists | Create a schema and or an API for accessing event list data. Activate the IHDEA working group on this topic and get people involved who have been curating event lists, and come up with a common representation for the many kinds of events within Heliophysics data, including future events from forecasts. This is a multi-year effort: collect use cases and examples, see what is available to leverage, develop standards, get buy-in, spread the word at meetings. | 1 | 5.00 | 3 | 15 |
| 16 | Common data products feasible? | Ask community what std data products should exist, eg SDO's plate-scaled/rectified images for ML. Is data 98% good 'good enough' or do you need to know the full processing chain? (eg morphology/forecasting vs photometry, etc) | 1 | 3.50 | 4 | 14 |
| 21 | Focus on 'commons' as the needed response to HDRL problems | Several of the challenges HDRL (and Helio writ large) face are perhaps as much social/cultural challenges as technical. Are we appropriately directing our work toward that? Perhaps we need to place our focus more fully on building the knowledge commons in earnest (treating heliophysics knowledge as a common resource that needs to be stewarded as such). A knowledge commons is not simply a collection of semantic artifacts. It is the integration of intelligent information representation with openness, governance, and community trust—an ecosystem in which the Heliophysics community collectively maintains and evolves shared conceptual infrastructure. | 1 | 4.33 | 3 | 13 |
| 41 | What will HDRL's minimum metadata be? | SPASE will have it's own minimal metadata requirements that will likely be much less than what HDRL needs. We need to discuss what the minimum requirements will be for all submitted data (mission and research). | 1 | 3.25 | 4 | 13 |
| 31 | What streamlining or combining can we do for our data curation processes? | Are there places in our metadata and data curation processes where we can increase efficiency? Are there easy things we can pursue to decrease the complexity of our systems? Please comment with ideas here or add them as separate ideas on the padlet. | 4 | 3.67 | 3 | 11 |
| 53 | Create a mechanism that would trivialize the collection process for observation and model data related to space weather events of interest | Any time there is a geomagnetic storm or astounding Coronal Mass Ejection (CME), etc, there is a lot of interest in tracing this event through from the Sun to the Earth. Build this in a way that each iteration provides some new and useful capability to at least one group of scientists, and then keep making it more useful over time. | 4 | 3.67 | 3 | 11 |
| 46 | Summarize SPASE records with AI | Purpose is to increase discovery of Helio datasets by AI searches. Idea is to summarize SPASE records with an AI tool, feed that summary to what the AI search tools use and see if the dataset is more highly prioritized for a given search. | 2 | 5.00 | 2 | 10 |
| 28 | Conjunction notebooks on the cloud | Add a 'Run on the Cloud' button on the magnetic conjunction notebooks to seamlessly let researchers run the magnetic conjunction notebooks on HSD's HelioCloud. How to get these pulled into the HelioData website? | 3, 5 | 2.00 | 4 | 8 |
| 45 | Copy NASA data served externally to here | Out of the ~3300 SPASE records, only about 2200 are for data we can assign the required license to (cc0). For the data that is NASA HPD funded, we need to copy the data for the remaining records to either SPDF or SDAC. We should contact the people at these other places to ask if there any any objections to the license. If there is no response after 2-4 weeks, we should move ahead to assign the cc0 license and serve the data from HDRL. | 5, 7 | 1.75 | 4 | 7 |
| 47 | Build RAG server for mission and researcher contributed documentation / Data | This should allow it to be hooked into helio.data and our services. | 3, 2 | 2.33 | 3 | 7 |
| 12 | Communicate uncertainties of data across domains | Outsiders have better understandings of data and services. HelioSystems Lab should include counter-factual use of data and enable discoveries. | 1 | 3.00 | 2 | 6 |
| **44** | **How to determine what data supported scientific breakthroughs?** | **Motivation here is to determine what data were used to achieve scientific breakthroughs. Need to start with determining what publications or other items indicated scientific breakthroughs, then which data were used as hopefully what services helped. Could determine the publications by popular vote on Helionauts?** | **6** | **1.67** | **3** | **5** |
| 55 | Make MCP servers for major HDRL services. | Make an MCP server which chooses relevant HDRL MCP server. Ask stuff and get automatically directed to the relevant service with transparency re: calls that are made. | 5 | 5.00 | 1 | 5 |
| 24 | ISTP standards for netCDF4/HDF5 | Need to establish what the standard format should be to express ISTP metadata in newer formats, including netCDF4/HDF5 | 1 | 1.00 | 2 | 2 |
| 49 | Understanding data access statistics | Problem is to produce reasonably accurate reporting of the number of data users, but bots are now in the statistics. Can we use AI to identify trends of concern to help inform blocking strategy? Perhaps pattern-based blocking? Could test with the log files. | 6 | 1.00 | 2 | 2 |

**Appendix C. Participant Feedback**

Google forms responses:
https://docs.google.com/forms/d/1_sG-pSVo9nNT2WCpgmBu0BA4JHkzkvOLTLtKfK1vN3o/edit#responses

Approximately ⅓ of the participants responded to a short feedback survey of the meeting (comprising 4 questions). A summary of their responses are given below. Note that the text responses were summarized using Gemini AI and reviewed/edited by the compilers of this report.

Q1. What do you think went well and/or was valuable for you?

- **Networking and Social Aspect:** Meeting and networking in person with other HDRL people, including core developers, and putting names to faces was highly valued.
- **Informative Content and Updates:** Learning about available services, work, different projects and their status, as well as future directions and short overview/updates on current HDRL-funded activities was extremely valuable and informative. The 3-minute show-and-tell format was particularly useful for understanding the program's facets.
- **Interactive Sessions and Engagement:** The group breakout sessions, particularly those using Padlet comments to generate ideas and thoughts, were effective and engaging. The variety of contributed ideas on Padlet and the subsequent discussions were appreciated.
- **Meeting Structure and Pace:** The pace of the meeting, with quick talks followed by questions, appropriate breaks, and the combination of 'Lightnings followed by Padlet, then rapid breakouts' was effective.
- **Purpose and Frequency:** The meeting was considered necessary for discussing what is going well/poorly and prioritizing new ideas, with a suggestion for increased frequency.

Q2. What do you think needs to be improved, changed or dropped from this meeting?

- **Discussion Format and Focus:** Discussions lacked focus, with time spent deciding what to discuss, and the plethora of inputs distracted from the meeting flow. Suggested improvements include longer breakout discussions, less repetition on the second day, and having some structured presentations.
- **Actionability and Assignments:** Concerns were raised about who will convert breakout outputs into reality and the need to assign groups/actions during the meeting to ensure follow-through, as without assignment, people will get bogged down by daily work. Breakouts should specifically ask for short-term and long-term actions and assignees.
- **Breakout Session Organization:** The first day's breakout sessions were disorganized and did not stay on target, though the second day was better. Less tech setup in the breakout rooms was also requested.

- **Packet Question Clarity:** Packet questions were sometimes interpreted differently and unclear; the person who had the idea should elaborate beforehand, especially if there is a breakout session.
- **Meeting Structure and Participants:** Consider adding another day for future work discussions, incorporating more overall strategy talks from organizers, and including more scientists in the discussions. A long lunch break was also mentioned.
- **Padlet Voting:** Suggestions were made to move Padlet voting "offline" to allow for more discussion time.

Q3. Meeting Rating

**Figure C1. Participant meeting rating**

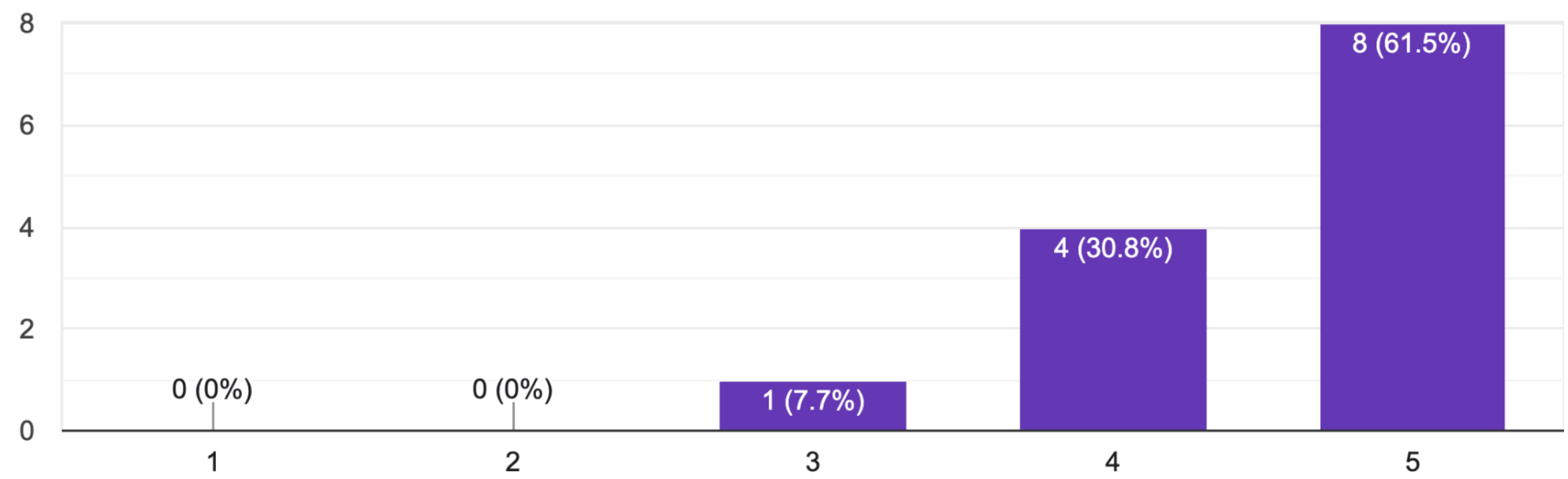

Q4. Any other feedback you care to share?

- **Meeting structure and agenda:** Suggestions included being more flexible in the schedule, the possibility of having too many open topics, and the need for a more frequent cadence (at least every two years, with one respondent looking forward to more frequent meetings).
- **Recap and follow-up:** A request was made to create an "as flown" recap from the breakouts, as the final recap was lossy, had fewer voices, and happened when people were tired. There was also concern about how to implement big changes when individuals have small roles and other issues lack dedicated working groups like SPASE.
- **Logistical issues:** Feedback highlighted the need for better WiFi support due to the online components of the meeting and concern was expressed about where to host the next meeting if building 21 is unavailable.

- **Positive aspects:** The meeting was considered valuable for staying informed about HDRL's work (presentations on WHAT HDRL is were helpful) and for connecting with colleagues, especially since many couldn't attend larger conferences.
- **Tools usability:** The Padlet was hard to follow during the voting period due to shuffling as new ideas entered, and a configuration update mode was suggested.

## Appendix D. HDMC Debrief Meeting

### Summary

A short meeting among some of the thought leaders in HDMC who attended the meeting, the 'glue' part of HDRL, was held on the day following the wider HDRL Strategy meeting. A summary of the notes are included here and then analyzed to compare the topics discussed to the trends noted this far.

The current convention of HDRL staff chasing down NASA-funded missions for delivery of data products was recognized as a significant drain of resources. Proposed solutions included new policies requiring future missions to acquire approval from HDRL for data management plans, requiring current missions to deliver the data to HDRL to ensure eligibility for future funding, and requiring future missions to deliver the data to HDRL once initial processing is completed. It was recognized that HDRL can quickly be ready to become the primary location of all NASA Heliophysics mission data, but needs more planning before data on the responsibility of hosting mission data processing.

Another recognized drain on HDRL resources was the high time-cost for adding events into current services, particularly the Heliophysics Event Knowledgebase (HEK). The proposed solution, although time-intensive, was to create a standard format for events. The effort was suggested as an IHDEA working group with staff from the HEK and related efforts to be tapped as potential leaders.

Creating SPASE records was a third recognized drain on staff time. The group expressed interest in the likely significant accelerations in this with the incorporation of AI/ML methods. Additionally, current AI/ML efforts, particularly the data-literature linking and the MCP work, were recognized as more valuable than previously thought due to the newly discussed capabilities they will bring for both staff and researcher usage.

Similarly, the attendees recognized a critical need for multiple taxonomies (e.g., phenomena, regions, parameter/variable types) and a dictionary for Heliophysics. These semantic artifacts were considered vital to support search and discovery not only of data, but of other resource types in a typical research workflow, including the typical examples publications and software, but also including people, conferences, and so on. Connecting Heliophysics resources together with these semantic artifacts as the standard language was seen as the next step in HDRL capabilities, additionally requiring continued collaboration with the Science Explorer.

The discussion on outreach was lively and covered several related topics, all seen as needed components of a healthily balanced outreach plan for HDRL. The components discussed included:

- HDRL workshops,
- Technical and user working groups,
- Focused demos and talks and science and mission meetings,
- The use of Helionauts for general feedback for HDRL, potentially posting AI-generated summaries of selected tutorials or meetings, and
- Carefully planned headlines and blogs for current HDRL projects.

Attendees noted various needs for these items, such as selecting the correct intended audience, the right meetings for tutorials, the correct HDRL staff needed to support each task, and making the materials visually compelling or interactive. The private nature of Helionauts was mentioned as a challenge, namely that it is not publicly searchable and can take some time to sign up. The proposed solution was for HDRL to help HelioNauts improve in these areas.

HDMC's cloud strategy discussion touched on several topics, including

- The lack of CS support for completing the paperwork process required for implementing HelioCloud on NCCS for Heliophysics researchers,
- Further development needed to fully develop a hybrid implementation of HelioCloud,
- The need for an easy method for researchers to cite the software environment used,
- A desire to indicate on HelioData what listed data is also available on the cloud, potentially through the addition of a "cloud" access type indicator, and
- The impending data limit for HSD's cloud storage considering upcoming usage by Heliophysics missions.

Attendees were interested in quarterly virtual meetings similar to this one, with yearly in-person meetings patterned after the HDRL strategy meetings but adapted for a smaller group.

### Analysis

The topics discussed in the HDMC debrief meeting align with several topic ideas and the themes derived from them as shown in Table 3 below. The first column shows a shortened title for each topic discussed, column 2 shows a list of the related ideas ordered by decreasing interest score, followed by the average ratings for those ideas. The four ideas that were selected for discussion during the wider HDRL Strategy meeting are indicated in bold text in the "Related Ideas" column. These were (36) "Build a knowledge graph of HDRL data", (32) "Draft SPASE records from documents using AI", (51) "HDRL Workshops", and (50) "How do we prepare to serve big data?". Ratings of 4.5 or above are highlighted with bold text. The table concludes with a list of the associated themes for each idea, determined by assembling the themes associated to each related idea into one list.

**Table D1: Connections for HDMC-Discussed Topics**

| Topic | Related Ideas | Avg. Ratings | Themes |
|---|---|---|---|
| Mission data delivery | 27 | **5.0** | 5 |
| Event data support | 5, 8, 54, 53 | **4.5**, 2.8, **5.0**, 3.7 | 1, 2, 4 |
| AI/ML efforts | **36**, 26, 33, **32**, 47, 55 | **4.7**, 4.3, 4.3, 4.3, 2.3, **5.0** | All but 7 |
| Helio. semantic resources | **36**, 19, 21 | **4.7**, **4.6**, 4.3 | 1, 2, 6 |
| HDRL outreach | **51** | **4.6** | 7 |
| Cloud and hybrid strategy | 52, **50** | **5.0**, **4.6** | 5 |

The matching between the HDMC topics and the related topic ideas is limited to the ideas discussed, not a general categorization of the topic ideas that were related to the HDMC topic. For example, the “AI/ML efforts” topic only touched on the MCP work, data-literature linking, and SPASE record creation, so other topics ideas related to AI/ML are excluded from the analysis. However, ideas that would be made more attainable by the discussed topics are included, such as the “Build a knowledge graph of HDRL data” idea for the “Heliophysics semantic resources” topic. More details on these and the other topic ideas indicated in Table 3 are available in Appendix B.